# On the van Cittert - Zernike theorem for intensity correlations and its applications

Timur E. Gureyev,[1,2,3,4,*] Alexander Kozlov,[1] David M. Paganin,[2] Yakov I. Nesterets,[4,3] Frank De Hoog,[5] Harry M. Quiney[1]

[1]*ARC Centre of Excellence in Advanced Molecular Imaging, The University of Melbourne, Parkville, VIC 3010, Australia*
[2]*School of Physics and Astronomy, Monash University, Clayton, VIC 3800, Australia*
[3]*School of Science and Technology, University of New England, Armidale, NSW 2351, Australia*
[4]*Manufacturing, Commonwealth Scientific and Industrial Research Organisation, Clayton, VIC 3168, Australia*
[5]*Data61, Commonwealth Scientific and Industrial Research Organisation, Canberra, ACT 2601, Australia*
*\*Corresponding author: timur.gureyev@unimelb.edu.au*

**A reciprocal relationship between the autocovariance of the light intensity in the source plane and in the far-field detector plane is presented in a form analogous to the classical van Cittert - Zernike theorem, but involving intensity correlation functions. A "classical" version of the reciprocity relationship is considered first, based on the assumption of circular Gaussian statistics of the complex amplitudes in the source plane. The result is consistent with the theory of Hanbury Brown - Twiss interferometry, but it is shown to be also applicable to estimation of the source size or the spatial resolution of the detector from the noise power spectrum of flat-field images. An alternative version of the van Cittert - Zernike theorem for intensity correlations is then derived for a quantized electromagnetic beam in a coherent state, which leads to Poisson statistics for the intrinsic intensity of the beam.**

## 1. INTRODUCTION

The van Cittert - Zernike (vCZ) theorem is a well-known result of classical statistical optics [1]. A generalized form of this theorem is sometimes referred to as the reciprocity relationship between the intensity distribution and the degree of (spatial) coherence of a quasi-monochromatic beam in the source and the detector planes [2]. This result involves spatial correlation functions of the second order [2, 3], i.e. statistical averages of the product of complex amplitudes of an electromagnetic beam at two different points on a plane orthogonal to the optic axis. Correlation functions of the fourth order with respect to complex amplitudes, such as statistical averages of the product of optical intensities at two different points, are used in the theory of the Hanbury Brown - Twiss (HBT) interferometer [2, 3, 4]. In the present paper we consider an extension of the vCZ theorem to such intensity correlation functions. Much of the related material can be found at different places in well-known textbooks [1, 2]. However, it appears that an explicit form of this vCZ theorem for intensity correlations (we will call it the vCZ2 theorem) may not be widely known, particularly in the general polychromatic partially coherent case that we consider here. We discuss how this result can be applied not only to source size measurements with the help of the HBT interferometer, but also for estimating the source size or the width of the detector point-spread function (PSF) from noise power spectra in so-called flat-field images (i.e. images obtained without any objects between the source and the detector) collected in a plane orthogonal to the optic axis. Such an approach can be useful for simple measurements of spatial resolution of optical imaging setups and instruments, eliminating the need for special resolution masks, sharp ("knife") edges or interferometric measurements in some cases. It could be useful, for example, for measuring coherence properties of synchrotron and X-ray free-electron laser (XFEL) sources, which was recently studied e.g. in [5, 6] using the HBT approach.

We also consider how the classical theory of HBT interferometry can be modified if one assumes a Poisson illumination model instead of Gaussian sources. As an analogue of the Gaussian moments theorem for multivariate Poisson statistics does not seem to exist, we abandon classical statistical optics in favour of a quantum electrodynamics treatment in this case. It turns out that the intensity autocovariance of a quantized electromagnetic beam in a coherent state does have Poisson statistics, which is intrinsic to the field's intensity variations, rather than being a feature of the probabilistic photon detection. This allows us to derive an alternative form of the vCZ2 theorem for intensity autocovariance with spatially correlated Poisson statistics. The variance due to the photon shot noise in the detector is still present, of course, and is added to the intrinsic variance of the field's intensity.

## 2. INTENSITY AUTOCOVARIANCE IN THE CASE OF GAUSSIAN STATISTICS

Let a quasi-monochromatic partially-coherent extended source be located immediately upstream from the "source plane" $z = 0$ orthogonal to the optic axis $z$. We want to calculate first the autocovariance of light intensity $I_R(x)$ at two points, $x_1$ and $x_2$, in a distant "detector plane" $z = R$ downstream from the source:

$$C_R(x_1, x_2) = <I_R(x_1) I_R(x_2)> - <I_R(x_1)><I_R(x_2)>, \quad (1)$$

where the angular brackets denote ensemble average. In order to simplify the notation, we will work with the one-dimensional case, i.e. the transverse coordinate $x$ will be a scalar. Extension of the main results to the two-dimensional case is straightforward.

We assume that the complex amplitudes in the source plane have circular Gaussian statistics. This assumption allows one to use the Gaussian moments theorem [1] for the fourth order correlations of the complex amplitudes $U_0(x')$ of light in the source plane:

$$<U_0^*(x_1') U_0^*(x_2') U_0(x_3') U_0(x_4')> = J_0(x_1', x_3') J_0(x_2', x_4') \\ + J_0(x_2', x_3') J_0(x_1', x_4'), \quad (2)$$

where $J_0(x_i', x_j') = <U_0^*(x_i') U_0(x_j')>$ is the mutual intensity of light at two points, $x_i'$ and $x_j'$, in the source plane. Equation (2) applies, for example, to sources consisting of multiple independent radiators, such as thermal light sources [1,2].

Under the conditions of paraxial illumination, we can express each complex amplitude in the detector plane via the Fresnel diffraction integrals with the integration over the source plane:

$$U_R(x) = \frac{\exp(ikR)}{i\sqrt{\lambda R}} \int \exp\left(\frac{i\pi}{\lambda R}(x-x')^2\right) U_0(x') dx', \quad (3)$$

where $\lambda$ is the radiation wavelength and $k = 2\pi/\lambda$. Evaluating the ensemble average $<I_R(x_1) I_R(x_2)> = <U_R^*(x_1) U_R^*(x_2) U_R(x_1) U_R(x_2)>$ with the help of eqs. (2) and (3), one obtains (see e.g. [1, 2]):

$$<I_R(x_1) I_R(x_2)> = <I_R(x_1)><I_R(x_2)> + |J_R(x_1, x_2)|^2 \\ = <I_R(x_1)><I_R(x_2)> [1 + |g_R(x_1, x_2)|^2], \quad (4)$$

where $I_R(x) = |U_R(x)|^2$, $J_R(x_1, x_2) = <U_R^*(x_1) U_R(x_2)>$ and $g_R(x_1, x_2) = J_R(x_1, x_2) / \left(<I_R(x_1)><I_R(x_2)>\right)^{1/2}$ is the corresponding complex coherence factor. Equations (1) and (4) imply that the autocovariance of light intensity in the detector plane is equal to

$$C_R(x_1, x_2) = |J_R(x_1, x_2)|^2. \quad (5)$$

In particular, the variance of the intensity in the considered case is equal to the square of its mean value, $V_R(x) \equiv C_R(x, x) = <I_R(x)>^2$. The last result shows that, under the model adopted here, the light in the detector plane has super-Poisson statistics, consistent with that of polarized thermal light [1]. This fact is a consequence of the assumption made before eq.(2) about the complex amplitude in the source plane having circular Gaussian statistics.

Let us now consider a general ergodic polychromatic beam and a model detector measuring average intensity over a time duration $T$ in a single exposure, $I_{T,R}(x) = T^{-1} \int_{-T/2}^{T/2} |U_R(x,t)|^2 dt$. Usually a light detector would be expected to integrate the instantaneous intensity over the exposure time, rather than average it. However, it is trivial to implement the division by the exposure time after the exposure. We will see that this normalization leads to results that have a convenient and natural mathematical form. It can be shown (see Appendix) that in this case eq.(5) becomes:

$$\begin{aligned} &C_{T,R}(x_1, x_2) \\ &\equiv <I_{T,R}(x_1) I_{T,R}(x_2)> - <I_{T,R}(x_1)><I_{T,R}(x_2)> \\ &= \iint W_R(x_1, x_2, \nu_1) W_R^*(x_1, x_2, \nu_2) \mathrm{sinc}^2[\pi(\nu_1 - \nu_2)T] d\nu_1 d\nu_2. \end{aligned} \quad (6)$$

where $\nu = c/\lambda$ is the (temporal) frequency and $W_R(x_1, x_2, \nu)$ is the cross-spectral density of the beam in the detector plane.

Let us consider two limit cases.

1. When the exposure time $T$ is so short and/or the radiation spectrum width $\Delta \nu$ is so narrow that $\mathrm{sinc}^2[\pi(\nu_1 - \nu_2)T] \cong 1$ over the support of $W_R(x_1, x_2, \nu)$, we have

$$C_{T,R}(x_1, x_2) \cong |\int W_R(x_1, x_2, \nu) d\nu|^2. \quad (7)$$

This can be seen as the "temporal coherent limit", corresponding to the case when the exposure time $T$ is much shorter than the coherence time $T_c$ of the beam incident on the detector (the coherence time is reciprocal to the width of the temporal spectrum, see the exact definition below). This case may be applicable only to some experiments involving imaging with high-quality single-mode laser light or XFEL sources [1, 5, 6]. For cross-spectrally pure light, i.e. when $W_R(x_1, x_2, \nu) = J_R(x_1, x_2) s(\nu)$, with $s(\nu) \geq 0$ and $\int s(\nu) d\nu = 1$, it is straightforward to verify that eq.(7) transforms into eq.(5).

2. When the exposure time is so long and/or the spectrum width is so broad that $\mathrm{sinc}^2[\pi(\nu_1 - \nu_2)T] \cong \delta(\nu_1 - \nu_2)/T$ (see Appendix for details regarding this approximation) over the support of $W_R(x_1, x_2, \nu)$, we have

$$C_{T,R}(x_1, x_2) \cong (1/T) \int |W_R(x_1, x_2, \nu)|^2 d\nu. \quad (8)$$

Note that the basic fact that the covariance should tend to zero when the exposure time tends to infinity can be understood directly from the first line of eq.(6): the right-hand side tends to zero in this case because for ergodic processes $T^{-1} \int_{-T/2}^{T/2} |U_R(x,t)|^2 dt \xrightarrow[T \to \infty]{} <|U_R(x,t)|^2> = I_R(x)$ for any $t$. The case described by eq.(8) corresponds to the "temporal incoherent limit", i.e. the case when the exposure time $T$ is much longer than the coherence time $T_c$. For cross-spectrally pure light, eq.(8) becomes

$$C_{T,R}(x_1, x_2) \cong (T_c/T) |J_R(x_1, x_2)|^2, \quad (8')$$

where $T_c = \int s^2(\nu) d\nu$ is an exact definition of the coherence time for cross-spectrally pure light [2, 7]. This definition, although it may seem counter-intuitive on first glance, actually quite accurately quantifies the generic notion of the coherence time as a quantity reciprocal to the width of the spectrum [2, 7, 8] (note that the condition

$\int s(\nu)d\nu = 1$ has been implicitly used here; without it, the definition becomes $T_c = \int s^2(\nu)d\nu / \int s(\nu)d\nu$).

The autocovariance can be quite small in the "temporally incoherent" case corresponding to eqs.(8)-(8'). Consider for example that for non-filtered visible light the coherence time is of the order of $10^{-15}$ s (at wavelength 0.5 μm and monochromaticity $\Delta\lambda/\lambda \sim 1$, $T_c \sim 1/\Delta\nu = \lambda^2/(c\Delta\lambda) \cong 1.7 \times 10^{-15}$ s), while a typical exposure time could be a few milliseconds, meaning that $T_c/T \sim 10^{-12}$ in this case. Obviously, the absolute value of the autocovariance is actually irrelevant, the important requirement is that this value should be larger than the typical errors in the measurements. For a single exposure, this requirement leads to the condition of the form: $\delta I(x)/I(x) < T_c/T$, where $\delta I(x)$ is the error in intensity measurement. In a well-designed experiment aimed at measurements of $C_{T,R}(x_1, x_2)$, the dominant error will likely arise from photon shot noise in the intensity measured by the detector. Note that in eq.(6) we assumed an ideal (deterministic) detector response for simplicity, but in general the effect of photon-counting statistics cannot be ignored [1, 2]. However, if spatially uncorrelated photon-counting statistics is assumed as usual, the error due to shot noise in the measurement of $C_R(x_1, x_2)$ will tend to zero at $x_1 \neq x_2$ (at least when the distance between the two points is larger than the width of the detector PSF), provided that the number of registered exposures used for calculating the ensemble averages according to the definition of $C_{T,R}(x_1, x_2)$, eq.(6), is sufficiently large. A detailed analysis of behavior of typical errors and signal-to-noise ratio in intensity interferometry can be found in [1, 2].

As the "temporally incoherent" case described by eqs.(8) and (8') is applicable to a broader class of optical experiments, compared to the "temporally coherent" case described by eq.(7) (as, the exposure time is typically much longer than the coherence time), we will mostly concentrate on this case in the following parts of the paper.

## 3. VAN CITTERT - ZERNIKE THEOREM FOR INTENSITY CORRELATIONS

Equation (8) can serve as a basis for the method for measuring the source size using the HBT interferometer [1-4]. For that purpose, one also needs to know how the cross-spectral density in the detector plane is related to relevant properties of the light source. Let us now assume that the cross-spectral density in the source plane can be described by the quasi-homogeneous model [1, 2]:

$$W_0(x_1', x_2', \nu) = S_0\left(\frac{x_1' + x_2'}{2}, \nu\right) g_0(x_1' - x_2', \nu), \quad (9)$$

where $S_0(x, \nu)$ is the spectral density distribution in the source plane and $g_0(x, \nu)$ is the corresponding spectral degree of coherence. In particular, $W_0(x', x', \nu) = S_0(x', \nu)$, as $g_0(0, \nu) = 1$ by definition. It is also assumed that the distribution $S_0(x, \nu)$ is much wider than $g_0(x, \nu)$, and $S_0(x, \nu)$ is almost constant over distances comparable with the diameter of the support of $g_0(x, \nu)$. Then, according to the generalized vCZ theorem [1] (termed "reciprocity relations" in [2]), the cross-spectral density in the far-field plane $z = R$ is equal to:

$$W_R(x_1, x_2, \nu) = \frac{\exp(i\psi)}{\lambda R} \hat{S}_0\left(\frac{x_1 - x_2}{\lambda R}, \nu\right) \hat{g}_0\left(\frac{x_1 + x_2}{2\lambda R}, \nu\right), \quad (10)$$

provided that $R \gg (\Delta s)^2 / \lambda$, where $\Delta s$ is the source size, $\psi = \pi(x_1^2 - x_2^2)/(\lambda R)$ and the hat symbol denotes the Fourier transform with respect to the position variable, $\hat{f}(u, \nu) = \int \exp(-i2\pi u x) f(x, \nu) dx$. Substituting eq.(10) into eq.(8), one obtains:

$$C_{T,R}(x_1, x_2)$$
$$\cong T^{-1} \int \frac{1}{\lambda^2 R^2} \left|\hat{S}_0\left(\frac{x_1 - x_2}{\lambda R}, \nu\right)\right|^2 \left|\hat{g}_0\left(\frac{x_1 + x_2}{2\lambda R}, \nu\right)\right|^2 d\nu, \quad (11)$$

when the detector exposure time is much longer than the coherence time of the beam. From eqs.(9) and (8) with $R = 0$ it also follows, that the autocovariance in the source plane is equal to

$$C_{T,0}(x_1', x_2')$$
$$\cong T^{-1} \int \left|S_0\left(\frac{x_1' + x_2'}{2}, \nu\right)\right|^2 |g_0(x_1' - x_2', \nu)|^2 d\nu. \quad (12)$$

Therefore, the autocovariance of the beam's intensity in the source plane and in a far-field detector plane have "reciprocal" dependences on the degree of coherence and the intensity distribution of the source. This behavior is clearly similar to the classical vCZ theorem and the reciprocity relations expressed by eqs.(9) and (10) [1, 2]. Therefore, we suggest this result be termed the van Cittert - Zernike theorem for intensity correlations (or vCZ2 theorem for brevity). A corresponding result for the case when the exposure time is much shorter than the coherence time can be obtained using eq.(7).

When the source is almost completely incoherent, we have $g_0(x, \nu) \cong \delta(x)$ in eq.(9), and hence $\hat{g}_0(\xi, \nu) \cong 1$ in eq.(11). In this case, the intensity autocovariance is proportional to the integral over the spectrum of the square modulus of the Fourier transform of the source spectral density distribution. This relationship allows one to measure the size of a distant incoherent source (e.g. a star) using the HBT interferometer.

## 4. MEASURING SOURCE SIZE AND DETECTOR RESOLUTION FROM FLAT-FIELD IMAGES

An alternative method to the HBT interferometer for measuring the source size or the spatial resolution of a detector system on the basis of eq.(11) can be implemented using so-called flat-field images, i.e. images collected by a position-sensitive (e.g. a CCD based) detector in the absence of any objects between the source and the detector.

We assume that the autocovariance function in the detector plane $C_{T,R}(x_1, x_2)$ can be described by eq.(11) and is spatially quasi-stationary, i.e. the function $\hat{g}_0^2(x/(\lambda R), \nu)$, $x = (x_1 + x_2)/2$, is almost constant across the image or a sufficiently large patch of it (which means that the source is almost completely incoherent). Then, according to the Wiener-Khinchin theorem [1], the Fourier transform of $C_{T,R}(x_1, x_2)$ with respect to the variable $r = x_1 - x_2$ over such a "stationary patch" coincides with the power spectral density of the corresponding noise process, $\Delta I_{T,R}(x) = I_{T,R}(x) - \langle I_{T,R}(x) \rangle$, integrated over the energy spectrum. Note that the noise power spectral density can be measured directly from the flat-field images by subtracting the average pixel intensity value over the stationary patch of the image and calculating the square modulus of the Fourier transform of the resultant zero-mean signal. This way we can get direct access to the Fourier transform (over the $r$ variable) of the function $\int (\lambda R)^{-2} \hat{S}_0^2(r/(\lambda R_1),\nu) \hat{g}_0^2(x/(\lambda R),\nu) d\nu$, which will be proportional to the autocorrelation of the source spectral density $S_0(x,\nu)$. In reality, however, the noise power spectrum in flat-field images will be inevitably affected by the PSF of the detector. Therefore, one also needs to take into account the effect of the detector PSF on the estimation of the source size by this method.

It is convenient at this point to re-write eq.(11) as

$$C_{T,R}(x_1,x_2) \cong T^{-1} \int \sigma^2\left(\frac{x_1+x_2}{2},\nu\right) \frac{H(x_1-x_2,\nu)}{H(0,\nu)} d\nu, \quad (13)$$

where $\sigma^2(x,\nu) = |\hat{S}_0(0,\nu)|^2 |\hat{g}_0(x/(\lambda R),\nu)|^2 /(\lambda^2 R^2)$ is the (noise) variance and $H(r,\nu) = |\hat{S}_0(r/(\lambda R),\nu)|^2$ characterises the spatial correlation of noise. We investigate how the intensity autocovariance described by eq.(13) changes as a result of linear filtering. A convolution with a non-negative filter function $P(x)$ (which integrates to unity and represents the PSF of a detector) can be written as:

$$I_{P,T,R}(x) = (I_{T,R} * P)(x) = \int I_{T,R}(x-y) P(y) dy. \quad (14)$$

In the case of spatially quasi-stationary statistics, when the noise variance, $\sigma^2(x,\nu)$, is changing slowly compared to $P(x)$, the autocovariance of $I_{P,T,R}(x)$ can be expressed with the help of eqs.(13) and (14):

$$\begin{aligned}
&<\Delta I_{P,T,R}(x_1) \Delta I_{P,T,R}(x_2)> \\
&= \iint C_{T,R}(x_1-y_1, x_2-y_2) P(y_1) P(y_2) dy_1 dy_2 \\
&\cong T^{-1} \int \sigma^2\left(\frac{x_1+x_2}{2},\nu\right) \frac{(H*P_2)(x_1-x_2,\nu)}{H(0,\nu)} d\nu,
\end{aligned} \quad (15)$$

where $\Delta I_{P,T,R}(x) = I_{P,T,R}(x) - <I_{P,T,R}(x)>$ and $P_2(x) = \int P(x+y) P(y) dy$ is the autocorrelation of the filter function. Taking the inverse Fourier transform of eq.(15) with respect to $r = x_1 - x_2$ over a "stationary patch" around the point $x = (x_1 + x_2)/2$ as suggested above, we find that the corresponding noise power spectrum, $W_x(\xi)$, is proportional to the product of the Fourier transform of $H(r,\nu)$ and the square of the modulation transfer function (MTF) $|\hat{P}(\xi)|^2$:

$$\begin{aligned}
W_{T,R,x}(\xi) \\
\cong |\hat{P}(\xi)|^2 T^{-1} \int \frac{1}{\lambda R}\left|\hat{g}_0\left(\frac{x}{\lambda R},\nu\right)\right|^2 \Gamma_0(\lambda R \xi, \nu) d\nu,
\end{aligned} \quad (16)$$

where $\Gamma_0(x',\nu) = \int S_0(y',\nu) S_0(x'+y',\nu) dy'$ is the autocorrelation function of the source spectral density distribution. When the latter distribution factorizes like $S_0(x',\nu) = I_0(x') s(\nu)$, we get $\Gamma_0(x',\nu) = \Gamma_0(x') s^2(\nu)$, $\Gamma_0(x') = \int I_0(y') I_0(x'+y') dy'$. Substituting this into eq.(16), we obtain

$$\begin{aligned}
W_{T,R,x}(\xi) \\
\cong \Gamma_0(\lambda R \xi) |\hat{P}(\xi)|^2 T^{-1} \int \frac{1}{\lambda R}\left|\hat{g}_0\left(\frac{x}{\lambda R},\nu\right)\right|^2 s^2(\nu) d\nu.
\end{aligned} \quad (17)$$

Equation (17) may allow one, under certain circumstances, to measure the MTF of the detector or the autocorrelation of the source intensity distribution (or, more generally, a combination of the two). In particular, if the autocorrelation of the source intensity distribution (expressed as a function of $\lambda R \xi$) is much narrower than the detector MTF, then it will primarily determine the width of the noise power spectrum. Thus, in this situation one should be able to estimate the source size from flat-field images. In the opposite case, when the MTF of the detector is much narrower than the autocorrelation of the source intensity, one should be able to estimate the width of the detector PSF. Note that the noise power spectrum of the (uncorrelated) photon shot noise in the detector will generally be flat and, therefore, will not affect the estimation of the cut-off frequency in this method (the effect of the detector PSF on this spectrum is already taken into account in eq.(17)).

As a more specific example, let us consider a situation which can occur in synchrotron-based hard X-ray microscopy. Here one may have a wavelength $\lambda$ of the order of 1 Å and a propagation distance $R$ of the order of 100 m. As $\lambda R = 10^4$ μm$^2$ here, the far-field conditions will be fulfilled for source sizes $\Delta s$ sufficiently smaller than 100 μm, such that $(\Delta s)^2 << \lambda R$. As can be seen from eq.(17), in order for the width of the noise power spectrum to be primarily defined by the source size, rather than the detector PSF, the following condition should be fulfilled: $\xi_{max} \equiv \Delta s /(\lambda R) < 1/\Delta x$, where $\Delta x$ is the width of the detector PSF $P(x)$. In our example $\xi_{max} < 10^{-2}$ μm$^{-1}$, meaning that $\Delta x$ can be equal to 100 μm or less. Note that here we assumed for simplicity that $P(x)$ has a Gaussian shape, so that the width of its Fourier transform is $1/\Delta x$, however, other functional forms of the detector PSF could also be easily considered. Finally, in order for the width of the source autocorrelation function $f_0(\xi) = \Gamma_0(\lambda R \xi)$ to be measurable from flat-field images, the spatial resolution $\Delta \xi$ of the noise power spectrum measurements should be sufficiently fine, so that $\lambda R \Delta \xi < \Delta s$. As $\Delta \xi = 1/(M \Delta x)$, where $M \Delta x$ is the width ("aperture") of the flat-field image, the last condition becomes $M \Delta x > (\lambda R)/\Delta s$. Given that a typical X-ray detector has a width of its PSF equal to a couple of pixels or so, and hence $M$ will be of the order of $10^2$ - $10^3$, the last resolution condition will be easily fulfilled. Therefore, under the conditions considered in this example, measurements of the noise power spectra in flat-field images should allow one to estimate the size of the X-ray source.

If, instead of the far field, we consider the case of "near-Fresnel" imaging, where $\lambda R \leq (\Delta s)^2$, then the validity conditions of vCZ theorem will not be fulfilled. However, it is possible to demonstrate [8] that eqs.(13)-(15) can still be valid and, as a consequence, a modified form of eq.(17) can be valid too:

$$W_{T,R,x}(\xi) \cong T^{-1} \hat{H}(\xi) |\hat{P}(\xi)|^2 G(x,R), \quad (18)$$

where $H(r)$ describes the correlation properties of the intensity incident on the detector and $G(x,R)$ is a slowly varying function which is almost constant over distances comparable to the size of spatial support of $H(r)$. If the incident intensity is nearly uncorrelated (between different pixels of the detector), the width of the function $\hat{H}(\xi)$ will be typically much larger than that of the MTF of the detector, $|\hat{P}(\xi)|^2$. Therefore, in this case, measuring the noise power spectrum of flat-field images allows one to estimate the width of the PSF of the detector [9]. This method has been known for some time in the context of characterization of spatial resolution of detectors [10]. The corresponding functionality has been implemented, for example, in publicly available software [11].

## 5. IMPLICATIONS OF GAUSSIAN STATISTICS

In this section, we will limit our considerations to the quasi-monochromatic case and omit the frequency variable $\nu = \nu_0$ and the exposure time $T$ from all expressions for simplicity. Within the constraints assumed by the quasi-homogeneous source model above, the free-space propagation of the (spatial) correlation coefficient of intensities in the Fraunhofer region can be obtained from the "coherent" version of eq.(11) (which is based on eq.(7), instead of eq.(8)):

$$c^{(2,2)}(x_1, x_2) \equiv \frac{C_R(x_1, x_2)}{[C_R(x_1, x_1) C_R(x_2, x_2)]^{1/2}}$$

$$= \frac{\left|\hat{S}_0\left(\frac{x_1 - x_2}{\lambda R}\right)\right|^2 \left|\hat{g}_0\left(\frac{x_1 + x_2}{2\lambda R}\right)\right|^2}{\left|\hat{S}_0(0)\right|^2 \left|\hat{g}_0\left(\frac{x_1}{\lambda R}\right) \hat{g}_0\left(\frac{x_2}{\lambda R}\right)\right|} = |\gamma^{(1,1)}(x_1, x_2)|^2, \quad (19)$$

where

$$\gamma^{(1,1)}(x_1, x_2) \equiv \frac{J_R(x_1, x_2)}{<I_R(x_1)>^{1/2} <I_R(x_2)>^{1/2}} = c^{(1,1)}(x_1, x_2) \quad (20)$$

is the second order coherence factor, which coincides with the correlation coefficient of the complex amplitudes. In turn, the fourth order coherence factor corresponding to intensity correlations can be expressed as

$$\gamma^{(2,2)}(x_1, x_2) \equiv \frac{<I_R(x_1) I_R(x_2)>}{<I_R(x_1)><I_R(x_2)>}$$

$$= \frac{<I_R(x_1)><I_R(x_2)> + |J_R(x_1, x_2)|^2}{<I_R(x_1)><I_R(x_2)>} \quad (21)$$

$$= 1 + c^{(2,2)}(x_1, x_2).$$

This implies, in particular, that complete second order coherence, $|\gamma^{(1,1)}(x_1, x_2)| = 1$, corresponds to the maximum of the absolute value of the fourth order coherence factor, which also corresponds to a complete lack of fourth order coherence, as $\gamma^{(2,2)}(x_1, x_2) = 2$ in this case. In turn, the complete coherence of intensity distributions, $\gamma^{(2,2)}(x_1, x_2) = 1$, corresponds to the minimum of the fourth order coherence factor, and implies complete incoherence of the complex amplitudes, $\gamma^{(1,1)}(x_1, x_2) = 0$. Note also that the complete second order coherence, $|\gamma^{(1,1)}(x_1, x_2)| = 1$, corresponds to the perfect correlation of the complex amplitudes, $|c^{(1,1)}(x_1, x_2)| = 1$, while the complete coherence of intensity distributions, $\gamma^{(2,2)}(x_1, x_2) = 1$, corresponds to the complete absence of intensity correlations, $c^{(2,2)}(x_1, x_2) = 0$, as follows from eq.(21). Furthermore, for fields produced by small incoherent sources ($\hat{g}_0 \equiv 1$), the intensity correlation coefficient will be close to 1 over a broad area in a distant image plane, as the width of the function $\hat{S}_0(\Delta x / (\lambda R))$ increases with $R$. Consequently, while the transverse coherence length of complex amplitudes (second order coherence) increases upon free-space propagation (see e.g. the behaviour of the coherence as a function of $R$ in the vCZ theorem), the coherence length of intensity distributions (fourth order coherence) actually decreases, because the correlation length of intensities increases as a function of R according to the vCZ2 theorem, and for intensity it means decoherence. In particular, small incoherent sources look incoherent in the fourth order (i.e. with respect to intensity), when viewed from a long distance.

The above results seem to contradict some intuitive expectations about the behavior of partially-coherent light fields. In the next section, we will use a quantum electrodynamics (QED) approach in order to consider an alternative model which has intrinsic Poisson, rather than Gaussian, statistics. This "Poisson illumination" model corresponds to the case of a perfectly coherent beam, as opposed to beams with low temporal coherence mostly considered in the previous sections. We will show that the counter-intuitive features of intensity correlations mentioned above do not appear in the model studied in the next section.

## 6. A QUANTUM VERSION OF THE VCZ2 THEOREM FOR COHERENT STATES

Consider a linearly polarized quantized electromagnetic field operator $E = \sum_k E_k$, with

$$E_k \equiv E_k(x, t) = E_k^+(x) \exp(-i\omega_k t) + E_k^- \exp(i\omega_k t),$$

$E_k^-(x) = (E_k^+)^\dagger(x)$ and $E_k^+(x) = i(\hbar \omega_k / 2)^{1/2} u_k(x) a_k$. Here $\hbar$ is the reduced Planck constant, $\omega_k$ are the angular frequencies of the modes, $u_k(x)$ are the mode functions, $a_k$ and $a_k^\dagger$ are the photon annihilation and creation operators, respectively [2, 3]. We assume the mode functions $u_k(x)$ to be orthonormal, in the usual sense that $\int dx \, u_{k_1}^*(x) u_{k_2}(x) = \delta_{k_1 k_2}$, where $\delta_{k_1 k_2}$ is the Kronecker symbol, but we do not limit the choice of these functions only to (truncated) plane waves, so the index $k$ here is just a label indexing a complete set of modes [3]. Let $|\alpha_k>$ be a (Glauber) coherent state corresponding to a single mode $k$ (for simplicity, we will not consider general multi-mode states here), labelled by a complex number $\alpha_k$. These states have photon numbers distributed according to Poisson statistics with the average number of photons equal to $\bar{n}_k = |\alpha_k|^2$. We will omit the single index "$k$" in the notation below for brevity, keeping in mind that we study a single-mode case.

A natural quantized analogue of the four-point autocorrelation of the complex amplitude used in eq.(2) is the expectation value

$$<\alpha | E_0^-(x_1') E_0^+(x_2') E_0^-(x_3') E_0^+(x_4') | \alpha> \\ = (\hbar \omega / 2)^2 (|\alpha|^4 + |\alpha|^2) u_0^*(x_1') u_0(x_2') u_0^*(x_3') u_0(x_4'), \quad (22)$$

where the right-hand side has been calculated using the standard properties of coherent states [2]. Here, like in eq.(2), the subscript index zero denotes that the functions are considered in the "source plane" $z = 0$. Note that, unlike eq.(2), the ordering of operators in the left-hand side of eq.(22) is important, as the operators $E_0^+$ and $E_0^-$ do not commute. The autocorrelation function in the form of eq.(19) is suitable for subsequent calculation of intensity correlations [12], as will also be seen below.

We can now calculate the four-fold Fresnel diffraction integral (according to eq.(3)) with respect to all the variables $x_1'$, $x_2'$, $x_3'$ and $x_4'$. This operation propagates equation (22) to the detector plane $z=R$. As we are interested in intensity correlations, we only write down the result for the case $x_1 = x_2$, $x_3 = x_4$ and then change the notation $x_3$ to $x_2$, to obtain:

$$< \alpha | G_R(x_1) G_R(x_2) | \alpha > \\ = (\hbar \omega / 2)^2 (|\alpha|^4 + |\alpha|^2) |u_R(x_1)|^2 |u_R(x_2)|^2, \quad (23)$$

where $G_R(x) = E_R^-(x) E_R^+(x)$ is the intensity operator at point $x$ in the plane $z=R$ and the subscript $R$ at the mode functions denotes the Fresnel transform in accordance with eq.(3) with $\lambda = \omega / (2\pi c)$. Similar, but simpler, calculations for a single intensity operator give:

$$< \alpha | G_R(x) | \alpha > = (\hbar \omega / 2) |\alpha|^2 |u_R(x)|^2, \quad (24)$$

leading to the following analogue of the autocovariance from eq.(4):

$$C_{\alpha, R}(x_1, x_2) \equiv < \alpha | G_R(x_1) G_R(x_2) | \alpha > \\ - < \alpha | G_R(x_1) | \alpha > < \alpha | G_R(x_2) | \alpha > \quad (25) \\ = (\hbar \omega / 2)^2 |\alpha|^2 |u_R(x_1)|^2 |u_R(x_2)|^2 .$$

As expected in the case of Poisson photon statistics, here the autocovariance is proportional to the average number of photons $\bar{n} = |\alpha|^2$. In particular, setting $x_1 = x_2 = x$, we obtain the following expression for the variance:

$$V_{\alpha, R}(x) \equiv C_{\alpha, R}(x, x) = < \alpha | G_R^2(x) | \alpha > \\ - (< \alpha | G_R(x) | \alpha >)^2 = (\hbar \omega / 2)^2 |\alpha|^2 |u_R(x)|^4 . \quad (26)$$

As usual, if the propagation distance $R$ is sufficiently large, the Fresnel diffraction integrals in eq.(25) can be replaced by the Fourier transforms:

$$C_{\alpha, R}(x_1, x_2) = \frac{(\hbar \omega / 2)^2 |\alpha|^2}{(\lambda R)^2} \left| \hat{u}\left(\frac{x_1}{\lambda R}\right) \right|^2 \left| \hat{u}\left(\frac{x_2}{\lambda R}\right) \right|^2 . \quad (27)$$

Equation (27) represents an analogue of the vCZ2 theorem, eq.(11), in the case of coherent states (Poisson statistics of the field intensity). In the source plane we naturally get an expression reciprocal to eq.(27):

$$C_{\alpha, 0}(x_1', x_2') = (\hbar \omega / 2)^2 |\alpha|^2 |u_0(x_1')|^2 |u_0(x_2')|^2 . \quad (28)$$

Equations (27) and (28) can be expressed in terms of the functions of the difference and half-sum of coordinates $(x_1, x_2)$ and $(x_1', x_2')$, respectively, to make them look even more similar to eqs.(11)-(12).

Similar to the case of the classical HBT interferometer analysis based on the assumption of circular Gaussian statistics in the source plane considered above, here eq.(27) allows one to estimate the width of the intensity distribution in the source plane from measurements of intensity autocovariance in a far-field detector plane. Indeed, the width $\sigma$ of the function $f(h) = C_{\alpha, R}(0, h)$ in eq.(27) will be equal to the width of the square modulus of the Fourier transform of the mode function. If this function has an approximately Gaussian shape, then the width of the square modulus of the mode itself will be approximately equal to $\lambda R / \sigma$. It may be interesting to compare this result with the statements made in some previous publications (see e.g. [4]) that the HBT signal is equal to zero in the case of coherent illumination. The reason for the discrepancy between the two results will be explained below.

For comparison with the Gaussian statistics case, it will be interesting to look at the equations for the second and fourth order correlation and coherence properties of an electromagnetic beam in a coherent state $|\alpha>$. Consistent with the nature of a coherent state, the absolute value of the second order coherence factor and the correlation coefficient are always equal to unity:

$$|\gamma_\alpha^{(1,1)}(x_1, x_2)| = c_\alpha^{(1,1)}(x_1, x_2) = 1. \quad (29)$$

The result for the fourth order coherence is less straightforward. Indeed, if one calculates the fourth order coherence factor of a beam in a coherent state using the conventional definition of correlation functions based on normal ordering of photon creation and annihilation operators [2, 3], it leads to perfect coherence of the fourth order. In particular,

$$\gamma_\alpha^{(2,2)}(x_1, x_2) \\ = \frac{< \alpha | E^-(x_1) E^-(x_2) E^+(x_1) E^+(x_2) | \alpha >}{< \alpha | E^-(x_1) E^+(x_1) | \alpha > < \alpha | E^-(x_2) E^+(x_2) | \alpha >} \quad (30) \\ = \frac{(\hbar \omega / 2)^2 |\alpha|^4 |u(x_1)|^2 |u(x_2)|^2}{[(\hbar \omega / 2) |\alpha|^2 |u(x_1)|^2][(\hbar \omega / 2) |\alpha|^2 |u(x_2)|^2]} = 1.$$

The result in eq.(30) is again fully consistent with the notion of coherent states. However, this approach leads to both the autocovariance and the variance of intensity being identically zero (indeed, the autocovariance is equal precisely to the difference between the numerator and the denominator of eq.(30)). This is clearly a non-physical outcome, as acknowledged by Glauber himself [3]. A more detailed discussion of this issue can be found in [13]. On the other hand, from eq.(23) and (24), we immediately obtain:

$$\gamma_\alpha^{(2,2)}(x_1, x_2) = \frac{< \alpha | G(x_1) G(x_2) | \alpha >}{I_\alpha(x_1) I_\alpha(x_2)} = 1 + |\alpha|^{-2}, \quad (31)$$

where $G(x) = E^-(x) E^+(x)$ is the intensity operator, and its expectation value $I_\alpha(x) = < \alpha | G(x) | \alpha > = (\hbar \omega / 2) |\alpha|^2 |u(x)|^2$ is the average intensity of the beam at a given point. The right-hand side of eq.(31) obviously approaches one, when the average number of photons $\bar{n} = |\alpha|^2$ becomes large. However, here the noise is non-zero and is consistent with the Poisson statistics. The intensity correlation coefficient is now easy to calculate too: $c_\alpha^{(2,2)}(x_1, x_2) = \gamma_\alpha^{(2,2)}(x_1, x_2) - 1 = |\alpha|^{-2} \xrightarrow[\bar{n} \to \infty]{} 0$, i.e. the intensity oscillations at different points quickly become almost completely uncorrelated as the number of photons in the beam increases. This is consistent with the usual physical picture, where the detection of photons at different points in the detector plane is independent from each other. Note that this could not be the case for a beam with a very low number of photons. Indeed, if a single photon present in the beam is detected at one point in the detector plane, it makes the (conditional) detection probability at any other point equal

to zero (as the photons have been exhausted by the single detection event).

Finally, we have the following expression for the signal-to-noise ratio:

$$SNR_R^2(x) \equiv \frac{I_{\alpha,R}^2(x)}{V_{\alpha,R}(x)} = \frac{|\alpha|^4}{|\alpha|^2} = |\alpha|^2 = \bar{n}, \quad \textbf{(32)}$$

as expected in the case of Poisson statistics. Note that here the Poisson statistics is exhibited by the intrinsic variations of the beam intensity (which has been quantized from the start), rather than as a result of the probabilistic detection process [1,2]. When the detection statistics is also taken into account, the two variances add up [1], resulting in higher noise and lower SNR in imaging with beams in coherent states compared to the incoherent ones with the same average number of photons. Note that the variance of the intrinsic noise in the incoherent states can be obtained from eqs.(8) or (8'), implying that it is proportional to the ratio of $T_c/T$, which is usually very small for beams with low temporal coherence.

## 7. CONCLUSIONS

We presented and discussed two forms of the result, that we termed the van Cittert - Zernike theorem for intensity correlations, which extends the classical reciprocity relationships between the intensity distribution and the degree of coherence in the source and image planes to the autocorrelations of optical intensity. It is well-known [1, 2] that, when the complex amplitudes in the source plane satisfy joint circular Gaussian statistics, the autocovariance of intensity in the Fraunhofer region is equal to the square modulus of the corresponding autocorrelation function of the complex amplitudes (i.e. the mutual intensity). Combining this with the classical vCZ theorem immediately gives one the desired fourth order reciprocity relationship (vCZ2 theorem). We also consider the implications of this approach in the case of arbitrary polychromatic beams and energy-integrating detectors. However, the assumption of circular Gaussian statistics in combination with that of quasi-monochromaticity appears to lead to some counter-intuitive results about the free-space propagation of fourth order correlation functions, as discussed in Section 5. This provided a motivation for studying the free-space propagation of intensity correlations in the case of Poisson, rather than Gaussian, intrinsic statistics. Coherent states of quantum optics were used for this purpose, with the resultant "quantum" version of the vCZ2 theorem leading to Poisson statistics for the intensity autocovariance. While the vCZ2 theorem directly corresponds to the Hanbury Brown - Twiss interferometry technique, we also proposed an alternative method for measuring the size of a distant light source based on the same theorem. In this method, the source size and/or the detector PSF can be measured from the noise power spectrum in flat-field (featureless) images, which should be possible to implement with conventional CCD-based detectors.

**Acknowledgment**. We thank Prof. Keith Nugent, A/Prof. Brian Abbey, Prof. Les Allen and Dr. Andrew Martin for helpful discussions.

## APPENDIX

Here we derive a formula for covariance of finite-time exposures of polychromatic partially coherent light:

$$C_{T,R}(x_1, x_2) = <I_{T,R}(x_1)I_{T,R}(x_2)> - <I_{T,R}(x_1)><I_{T,R}(x_2)>, \quad \textbf{(A1)}$$

where $I_{T,R}(x) = T^{-1} \int_{-T/2}^{T/2} |U_R(x,t)|^2 \, dt$ is the instantaneous light intensity averaged over the time interval $T$. Expressing the complex amplitudes via the Fourier transforms over the time variable

$$U_R(x,t) = \int \tilde{U}_R(x,\nu) \exp(i2\pi\nu t) d\nu, \quad \textbf{(A2)}$$

and substituting the result into (A1), we obtain:

$$\begin{aligned}
&<I_{T,R}(x_1)I_{T,R}(x_2)> \\
&= <T^{-2} \int_{-T/2}^{T/2} |\int \tilde{U}_R(x_1,\nu_1)\exp(i2\pi\nu_1 t_1)d\nu_1|^2 \, dt_1 \\
&\times \int_{-T/2}^{T/2} |\int \tilde{U}_R(x_2,\nu_2)\exp(i2\pi\nu_2 t_2)d\nu_2|^2 \, dt_2> \\
&= \iiiint T^{-2} \int_{-T/2}^{T/2}\int_{-T/2}^{T/2} \exp[i2\pi(\nu_1-\nu_1')t_1]\exp[i2\pi \\
&\times (\nu_2-\nu_2')t_2] F_R(x_1,\nu_1,\nu_1',x_2,\nu_2,\nu_2') \, dt_1 dt_2 d\nu_1 d\nu_1' d\nu_2 d\nu_2' \\
&= \iiiint \operatorname{sinc}[\pi(\nu_1-\nu_1')T]\operatorname{sinc}[\pi(\nu_2-\nu_2')T] \\
&\times F_R(x_1,\nu_1,\nu_1',x_2,\nu_2,\nu_2') d\nu_1 d\nu_1' d\nu_2 d\nu_2',
\end{aligned} \quad \textbf{(A3)}$$

with

$$F_R(x_1, v_1, v_1', x_2, v_2, v_2')$$
$$= <\tilde{U}_R(x_1, v_1)\tilde{U}_R^*(x_1, v_1')\tilde{U}_R(x_2, v_2)\tilde{U}_R^*(x_2, v_2')>. \quad \text{(A4)}$$

Let us represent the linear transformation (propagator) of a complex amplitude from the source plane $z = 0$ to the detector plane $z = R$ formally as a linear operator $\tilde{U}_R = L_{R,v}[\tilde{U}_0]$. This operator can be interchanged with the ensemble average in eq.(A4):

$$F_R(x_1, v_1, v_1', x_2, v_2, v_2') = L_{R,x_1,v_1} L_{R,x_1,v_1'} L_{R,x_2,v_2} L_{R,x_2,v_2'}$$
$$<\tilde{U}_0(x_1', v_1)\tilde{U}_0^*(x_1', v_1')\tilde{U}_0(x_2', v_2)\tilde{U}_0^*(x_2', v_2')>. \quad \text{(A5)}$$

Now we apply the Gaussian moments theorem in the source plane:

$$F_R(x_1, v_1, v_1', x_2, v_2, v_2') = L_{R,x_1,v_1} L_{R,x_1,v_1'} L_{R,x_2,v_2} L_{R,x_2,v_2'}$$
$$[<\tilde{U}_0(x_1', v_1)\tilde{U}_0^*(x_1'', v_1')><\tilde{U}_0(x_2', v_2)\tilde{U}_0^*(x_2'', v_2')> \quad \text{(A6)}$$
$$+ <\tilde{U}_0(x_1', v_1)\tilde{U}_0^*(x_2'', v_2')><\tilde{U}_0(x_2', v_2)\tilde{U}_0^*(x_1'', v_1')>].$$

Taking the linear operators back inside the ensemble averaging, we now obtain:

$$F_R(x_1, v_1, v_1', x_2, v_2, v_2')$$
$$= <\tilde{U}_R(x_1, v_1)\tilde{U}_R^*(x_1, v_1')><\tilde{U}_R(x_2, v_2)\tilde{U}_R^*(x_2, v_2')>$$
$$+ <\tilde{U}_R(x_1, v_1)\tilde{U}_R^*(x_2, v_2')><\tilde{U}_R(x_2, v_2)\tilde{U}_R^*(x_1, v_1')>.$$

By definition, $<\tilde{U}_R(x_1, v)\tilde{U}_R^*(x_2, v')> = W_R(x_1, x_2, v)\delta(v - v')$ and $<\tilde{U}_R(x, v)\tilde{U}_R^*(x, v')> = S_R(x, v)\delta(v - v')$, where $W_R(x_1, x_2, v)$ is the cross-spectral density and $S_R(x, v) = W_R(x, x, v)$ is the corresponding spectral density. Therefore,

$$F_R(x_1, v_1, v_1', x_2, v_2, v_2')$$
$$= S_R(x_1, v_1)\delta(v_1 - v_1')S_R(x_2, v_2)\delta(v_2 - v_2') \quad \text{(A7)}$$
$$+ W_R(x_1, x_2, v_1)\delta(v_1 - v_2')W_R^*(x_1, x_2, v_2)\delta(v_2 - v_1').$$

Substituting (A7) into (A3) we obtain:

$$<I_{T,R}(x_1)I_{T,R}(x_2)> =$$
$$= \int S_R(x_1, v_1)dv_1 \int S_R(x_2, v_2)dv_2 + \iint W_R(x_1, x_2, v_1) \quad \text{(A8)}$$
$$\times W_R^*(x_1, x_2, v_2)\text{sinc}^2[\pi(v_1 - v_2)T]dv_1 dv_2.$$

Note that $\int S_R(x, v)dv = <I_{\infty,R}(x)> = <I_{T,R}(x)>$ for any $T$, as the ensemble average is time-independent, assuming that the process is wide-sense stationary. Therefore, the covariance is equal to

$$C_{T,R}(x_1, x_2) = \iint W_R(x_1, x_2, v_1)$$
$$\times W_R^*(x_1, x_2, v_2)\text{sinc}^2[\pi(v_1 - v_2)T]dv_1 dv_2. \quad \text{(A9)}$$

When $T \to 0$, it is easy to see that $\text{sinc}^2[\pi vT] \cong 1$ for any fixed $v$ (note that in any real optical system, $v$ cannot be arbitrarily large). On the other hand, when $T \to \infty$, $\text{sinc}^2[\pi vT] \cong (1/T)\delta(v)$. The last approximation can be demonstrated as follows. Consider the function $f_T(x) = \sin^2(\pi vT)/(\pi^2 v^2 T)$, such that $\text{sinc}^2(\pi vT) = (1/T)f_T(v)$. Firstly, note that

$$f_T(v) \xrightarrow[T \to \infty]{} \begin{cases} \infty, & \text{if } v = 0, \\ 0, & \text{if } v \neq 0. \end{cases} \quad \text{(A10)}$$

Indeed, when $v \to 0$ for any fixed $T$, we have (expanding sine function into the Taylor series) $f_T(v) = T - (2/3)\pi^2 v^2 T^3 + O(T^5) \xrightarrow[T \to \infty]{} \infty$. On the other hand, if $v \neq 0$ and is fixed, then, when $T \to \infty$, the numerator of $\sin^2(\pi vT)/(\pi^2 v^2 T)$ cannot exceed 1, while the denominator tends to infinity, so the fraction tends to zero.

Secondly, note that

$$\int_{-\infty}^{+\infty} f_T(v)dv = 1 \text{ for any T.} \quad \text{(A11)}$$

Here is the proof:

$$\int_{-\infty}^{\infty} f_T(v)dv = \int_{-\infty}^{\infty} \frac{\sin^2(\pi vT)}{\pi^2 v^2 T}dx$$
$$= \frac{1}{2\pi^2 T}\int_{-\infty}^{\infty} \frac{1 - \cos(2\pi vT)}{v^2}dv$$
$$= -\frac{1}{2\pi^2 T}\frac{1 - \cos(2\pi vT)}{v}\bigg|_{-\infty}^{+\infty} + \frac{1}{\pi}\int_{-\infty}^{\infty} \frac{\sin(2\pi vT)}{v}dv$$
$$= \frac{1}{\pi}\int_{-\infty}^{\infty} \frac{\sin(v)}{v}dv = 1 \text{ for any } T.$$

It is known (and is easy to prove) that any function $f_T(v)$, which satisfies eqs.(A10) and (A11), has the following limit: $f_T(v) \xrightarrow[T \to \infty]{} \delta(v)$, in a rigorous sense that $\int f_T(v)g(v)dv \xrightarrow[T \to \infty]{} g(0)$ for any smooth function $g(v)$ with a compact support.